\documentclass{article}

\usepackage{arxiv}

\usepackage[utf8]{inputenc} 
\usepackage[T1]{fontenc}    
\usepackage{hyperref}       
\usepackage{url}            
\usepackage{booktabs}       
\usepackage{amsfonts}       
\usepackage{nicefrac}       
\usepackage{microtype}      
\usepackage{lipsum}		
\usepackage{graphicx}
\usepackage{multirow}
\usepackage{subfigure}
\usepackage{hhline}
\usepackage{ltablex,booktabs}

\title{LV-Linker: Supporting Linked Exploration of Phone Usage Log Data and Screen Video Data}

\author{
    Hansoo Lee \\
    School of Computing\\
    KAIST\\
    Daejeon, Republic of Korea\\
    \texttt{hansoo@kaist.ac.kr}
    \And
    Sangwook Lee\\
    School of Computing\\
    KAIST\\
    Daejeon, Republic of Korea\\
    \texttt{sangwooklee@kaist.ac.kr}
    \And
    Youngji Koh\\
    School of Computing\\
    KAIST\\
    Daejeon, Republic of Korea\\
    \texttt{youngji@kaist.ac.kr>}\
    \And
    Uichin Lee\\\
    School of Computing\\
    KAIST\\
    Daejeon, Republic of Korea\\
    \texttt{uclee@kaist.ac.kr}
    }

\hypersetup{
pdftitle={A template for the arxiv style},
pdfsubject={q-bio.NC, q-bio.QM},
pdfauthor={David S.~Hippocampus, Elias D.~Striatum},
pdfkeywords={First keyword, Second keyword, More},
}

\begin{document}
\maketitle

\begin{abstract}
Prior HCI studies often analyzed smartphone app usage data for usability and user experience research purposes. App usage videos are often collected by a screen recording app in order to better analyze users’ app usage behaviors (e.g., app usage time, screen transition, and notification handling). However, it is difficult to analyze app usage videos along with multiple user interaction stream data. When the length of a video is long, data analysis tends to take a long time due to the volume of user interaction data. This is even more difficult for novice researchers due to a lack of data analysis experience. In this paper, we propose LV-Linker (Log and Video Linker), a visualization tool that helps researchers quickly explore the app usage log and video data by linking multiple time series log data with the video data. We conducted a preliminary user study with eight participants to evaluate the benefits of linking, by measuring task completion time, helpfulness, and subjective task workload. Our results showed that offering a linking feature significantly lowers the task completion time and task workload.
\end{abstract}

\section{Introduction}
Smartphone app usage data collected through user-smartphone interaction is utilized in various industries and HCI research fields. Representing, UI/UX designers use app usage data-driven UI/UX design for smartphone home screen and app UI/UX enhancement. Furthermore, the app usage data is being used in many user-centric computing research fields (e.g., digital phenotype, understanding user's smartphone usage pattern, improving user's interaction \& accessibility, enhancement of UI/UX) \cite{boukhechba2017monitoring, church2015understanding, lee2014hooked, singh2016cooperative, stutz2015smartphone, chen2019messageontap, riegler2015ui, zhang2017interaction}. 

The recorded app usage video data can be used to understand the user's app usage behavior patterns. There are two ways to collect app usage video data: (1) using a smartphone system built-in screen recording app or third-party screen recording app, and (2) collecting through a wearable camera. However, in the case of Android and iOS, which are representative smartphone OS, Android 10 released in September 2019 and iOS 11 released in June 2017 supported their screen recording function, respectively \cite{Android10, iOS11}. Therefore, previous studies published before 2019 identified app usage behavior patterns by collecting app usage video data through an independently developed screen recording system. When recording the surrounding context as well as the smartphone screen to analyze the user's app usage behavior pattern that changes by the ambient environmental factors, existing studies used a wearable camera to capture the surrounding context as well as the smartphone screen \cite{brown2013iphone, heitmayer2021smartphones}.

This app usage video data shows detailed app usage information to identify the user's app usage behavior pattern and can be easily used by researchers without learning difficulties. However, a lengthy video data is difficult to pinpoint a specific point of interest, and it takes a long time to analyze the multiple data streams.  Additionally, when collecting app usage video data as an in-the-wild study, not based on a pre-defined scenario, there are potential privacy issues because app usage videos  may capture personal information \cite{krieter2018analyzing}. It is necessary to remove the sensitive part of a video, but it is difficult to edit lengthy videos. In addition, exploring app usage video is difficult to accurately locate associated quantitative data (e.g., app usage time).

Existing studies use time-series app usage log data consisting of text, numerical, and categorical data to identify users’ app usage behavior patterns \cite{boukhechba2017monitoring, church2015understanding, lee2014hooked, stutz2015smartphone, chen2019messageontap}. These app usage log data can be collected by the app usage logger developed by Android/iOS built-in APIs provided for third-party app developers such as UsageStatsManager and AccessibilityService \cite{AccessibilityService, UsageStatsManager} in Android OS. Aware Framework is one of the representative app usage loggers for app usage analysis \cite{ferreira2015aware}. These app usage loggers can collect the app usage log data such as app usage status (foreground/background/stop time, app \& package name, screen transition), view hierarchy/elements, key typing, notification, and touch interaction type (e.g., click, long click, swipe, scroll). 

App usage log data has many advantages in terms of research to identify app usage behavior patterns compared to app usage video data. This app usage log data includes touch interaction information that occurs when a user interacts with a smartphone, which cannot be identified with app usage video data. In addition, app usage log data, which is time-series data of numerical, categorical, and text types of data, is easy to remove sensitive data through preprocessing. Therefore, using app usage log data for research is easier to protect privacy issues than using app usage video data. Furthermore, even if the app usage log data is collected for a long time, the capacity is relatively small compared to the app usage video data, and it is possible to quickly find and analyze a specific point in time through timestamp information on the spreadsheet. In addition, app usage log data has the advantage of being easier to perform data analysis (e.g., statistical analysis and machine learning) compared to app usage video data. Therefore, since app usage log data can accurately identify app usage behavior patterns in real-time, it is currently used in various research fields more than app usage video data to quantitatively analyze app usage behavior patterns \cite{boukhechba2017monitoring, church2015understanding, lee2014hooked}. 

However, to collect app usage log data, an app usage logger must be developed using Android built-in APIs, and there are various data types and attributes as well. Therefore, there is a learnability issue in collecting app usage log data and using it for research. In this study, we propose an LV-Linker which is Log and Video Linker, a viewpoint movement interaction system that synchronizes the time of app usage video and logs data, and moves the viewpoint of one data point as well as the viewpoint of another data. LV-Linker is a system configured by combining a spreadsheet-based log viewer for app log data and a video player-based video viewer for app video data. Through this research, we expect that the proposed system help researchers to analyze the app behavior patterns more easily, quickly, and accurately than when they used both types of data separately.

\section{Related Work}
Existing studies that connect time-series log and video focus on linking video and animation or text for audio description of the video \cite{mu2010towards, yamamoto2008video, merkt2011learning, kwan2018development}. These studies have in common with our study in that they can move the viewpoint through the interaction between video and text to analyze or find important viewpoints with annotations or text-connected timestamps. However, those studies are different from our study in that they did not use time series-based app usage log data, and they are not targeting a mobile environment.

The previous study in a mobile environment is to create a system replayer that can summarize statistical data or focus on specific factors of interest by synchronizing images taken while using mobile systems and logs generated from the system \cite{morrison2006coordinated}, and there is a systematic study that allows users to check wearable cameras in which surrounding environments are recorded for user experience recording when using smartphones \cite{brown2013iphone}. However, in existing studies, the video was not filmed on a mobile screen, but filmed from afar with a camera, and rather than focusing on the connection with app usage data, it was designed to view multiple people's mobile system logs and usage videos at the same time and check the user's context. Therefore, this study differs from previous studies in that it focuses on connecting recorded videos using Android app usage logs and recorded apps inside smartphones.

Therefore, this study proposes LV-Linker, a system that connects screen recording video data and time series app log data that can support app usage analysis in terms of data understanding, correctness, and quickness in a mobile environment. Accordingly, the purpose of this study to understand the linking effect of app usage log and video data is as follows: (1) We compare the performance of the proposed system and app logs/recorded videos in terms of exploring app usage data. (2) We examine how the prior experience of analyzing app usage affects the use of the proposed system compared to app logs and recorded video. (3) We examine how LV-Linker helps to perform app usage data analysis tasks compared to app logs and recorded videos in terms of correctness, quickness, and data understanding.

\begin{figure}
  \centering
  \includegraphics[width=\textwidth]{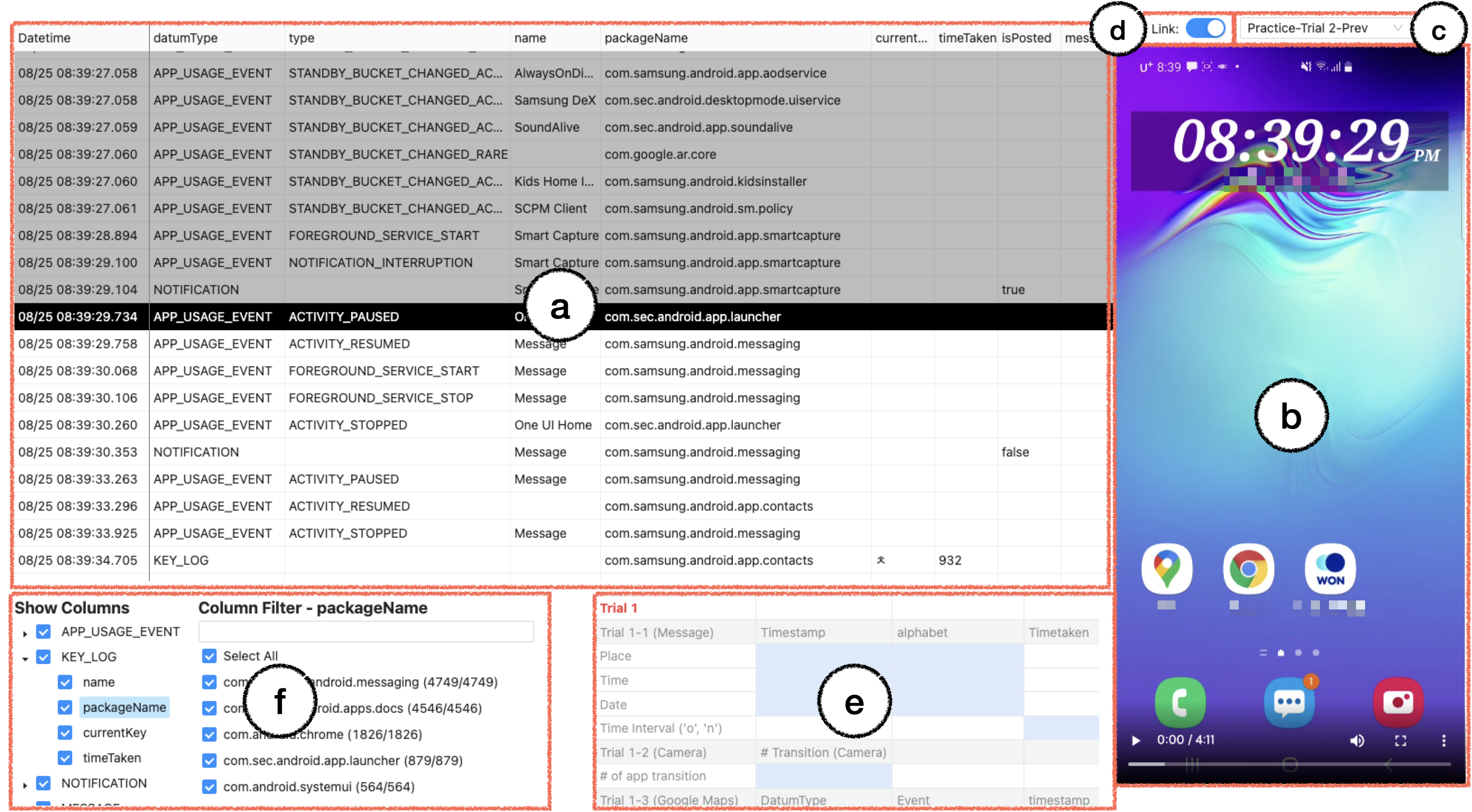}
  \caption{The user interface of LV-Linker: (a) log viewer, (b) video player, (c) dropdown for video selection, (d) link/unlink toggle switch, (e) task sheet, (f) filter function}
  \label{system}
\end{figure}

\section{LV-Linker: Linking Usage Log \& Video}
\subsection{System Description}
We developed LV-Linker to help researchers evaluate how linking app usage logs and videos affects users' correctness, quickness, data understanding, and experience when analyzing smartphone app usage data. Log viewer shows the processed logs in the form of a spreadsheet as shown in Figure \ref{system} \textbf{(a)}. The video player shows the app usage video that records users’ app usage behavior pattern on the smartphone in Figure \ref{system} \textbf{(b)}. It supports a basic seek bar to change into a specific frame. The users can select the video with the dropdown menu as depicted in Figure \ref{system} \textbf{(c)}. LV-Linker integrates the log viewer that shows the processed logs in a spreadsheet into the video player that shows user behavior on the smartphone as depicted in Figure \ref{system}. We added a toggle switch to measure the effects of linking only to link or unlink the log and video in the same UI as shown in Figure \ref{system} \textbf{(d)}. Furthermore, the system embedded the task sheet for participants to immediately copy and paste the required analysis results as depicted in Figure \ref{system} \textbf{(e)}. The system was implemented and deployed on the web frontend based on TypeScript and React.

\subsection{LV-Linker Usage Scenarios and Interaction Examples for Evaluation}
To showcase the benefits of LV-Linker, we consider the following data analysis tasks that could be used in the usability evaluation experiments. Twelve videos were recorded by a smartphone screen recording app at different times. Each video contains 6-minute app usage behaviors of the following five app tasks which are the most frequently used smartphone apps \cite{Statista:mostfrequentlyusedapps, Statista:mostusedsmartphonefunctions}. For the following smartphone app tasks, we created and recorded data on our own, such as making calls and sending text messages using two smartphones using virtual names.

\begin{itemize}
\item Answering a call: Leave a different message depending on the person who called.
\item Sending SMS: Send a reply to a message that makes an appointment for a time, place, and date.
\item Taking pictures: Take pictures of business cards in the camera app and delete one of them in the gallery app.
\item Sharing a route: Find a route to a specific store in Google Maps and share the route link through a message.
\item Transferring money: Transferring money to the account provided in a message in a banking app. 
\end{itemize}

We ordered these videos randomly and placed them within the dropdown menu so that participants can move on to the following video for every task as depicted in Figure \ref{system} \textbf{(c)}.

We used a self-developed app usage logger by Android built-in APIs \cite{AccessibilityService, UsageStatsManager, NotificationListenerService, NotificationManager} to collect the app usage logs such as the app usage status (e.g., start \& end time, app \& package name), notification (e.g., posted, app \& package name), and typing event (typed letters \& time, app \& package name), touch interaction type (e.g., click), device event (screen on/off, battery status, power on/off), call \& SMS logs. The app usage logs consist of metadata such as timestamp, datum type, and JSON data with various attributes (e.g., package name, app name, type, is posted). Our system appends metadata to the parsed JSON data and presents the data on the sheet. 

As shown in Figure \ref{system} \textbf{(f)}, Users can select the data they want to see using the filter function on the Log viewer. The filter consists of two stages: (1) Show/Hide columns, (2) Filter by the value of the selected column. The columns are grouped by datumType, which is the type of app usage log data (e.g., app usage event, notification, key log, device event) because each datumType has different sets of columns. To simplify the task for the experiment, we used a filter feature to exclude useless columns and datum types (e.g., device event, call \& SMS, touch interaction type). The remaining logs belong to APP\_USAGE\_EVENT, KEY\_LOG, and NOTIFICATION datumType. The remaining columns are timestamp, datumType, type (e.g., app usage interaction type), name, package name, currentKey, timeTaken, isPosted. Table \ref{appusagelog} has detailed descriptions and data types about datumType and columns of app usage log data used in LV-Linker.

\begin{table}[]
\caption{Description of filtered app usage log data used for usability evaluation experiments of LV-Linker System}
\label{appusagelog}
\resizebox{\textwidth}{!}{
\begin{tabular}{llll}

\hline
\multicolumn{1}{c}{\textbf{DatumType}} & \multicolumn{1}{c}{\textbf{Column}} & \multicolumn{1}{c}{\textbf{Data Type}} & \multicolumn{1}{c}{\textbf{Description}} \\ \hhline{====}

                                    &                        & Activity\_paused  & App moves to the background                                                      \\
                                    &                        & Activity\_resumed & App moves to the foreground                                                      \\
                                    & \multirow{-3}{*}{Type} & Activity\_stopped & App activity is terminated                                                       \\
                                    & Name                   & string            & User expression name of the app in the current event state                       \\ 
                                    & packageName            & string            & Developer package name of the app in the current event state                     \\
\multirow{-6}{*}{APP\_USAGE\_EVENT} & timestamp              & numerical         & Timestamp for the current event state                                            \\ \hline
                                    & currentKey             & string            & The currently entered key raw value                                              \\
                                    & timeTaken              & numerial          & Time difference between prevKey and currentKey input                             \\
                                    & Name                   & string            & User expression name of the app in the current key value \\
                                    & packageName            & string            & Developer package name of the app in the current key value                       \\
\multirow{-5}{*}{KEY\_LOG}          & timestamp              & numerical         & Timestamp for the current key value                                              \\ \hline
                                    & isposted               & boolean           & Whether the notificationatin is currently posted or not                          \\
                                    & Name                   & string            & User expression name of the app in the current notification event                \\
                                    & packageName            & string            & Developer package name of the app in the current notification event              \\
\multirow{-4}{*}{NOTIFICATION}      & timestamp              & numerical         & Timestamp for the current notification event                                     \\ \hline

\end{tabular}
}
\end{table}
The app usage log and video data are synchronized based on the timestamp. Thus, when users select a specific frame on the video player, the log viewer highlights a log recorded at the same frame. Conversely, when the users select a specific log on the log viewer, the video player seeks the corresponding time frame on the video player. Then, we manually synchronized the log and video based on one key typing event that we can check in both. Thus, there is a slight sync error between them.

\section{Preliminary Evaluation}
\subsection{Evaluation Setup and Procedure}
We recruited eight university students (5 female, 3 male) through online community posting. Through the preliminary survey, we divided participants into two groups; experienced and non-experienced people. Only two of them have prior experience in analyzing app usage data. The experiment was 90 minutes long. The research was exempt from the Institutional Review Board (IRB) under the condition that we do not collect  participants' personal information.

The experiment focused on exploring the effects of linking logs and videos when analyzing app usage data. Before experimenting, participants were asked to perform a preliminary survey. The preliminary survey asked questions about prior knowledge of Android app usage log data, and smartphone information (model name, OS) they are using. Through this survey, participants were classified into experienced groups and non-experienced groups according to prior knowledge of app usage data.

\begin{figure}
\centering    
\subfigure{\includegraphics[width=0.48\textwidth]{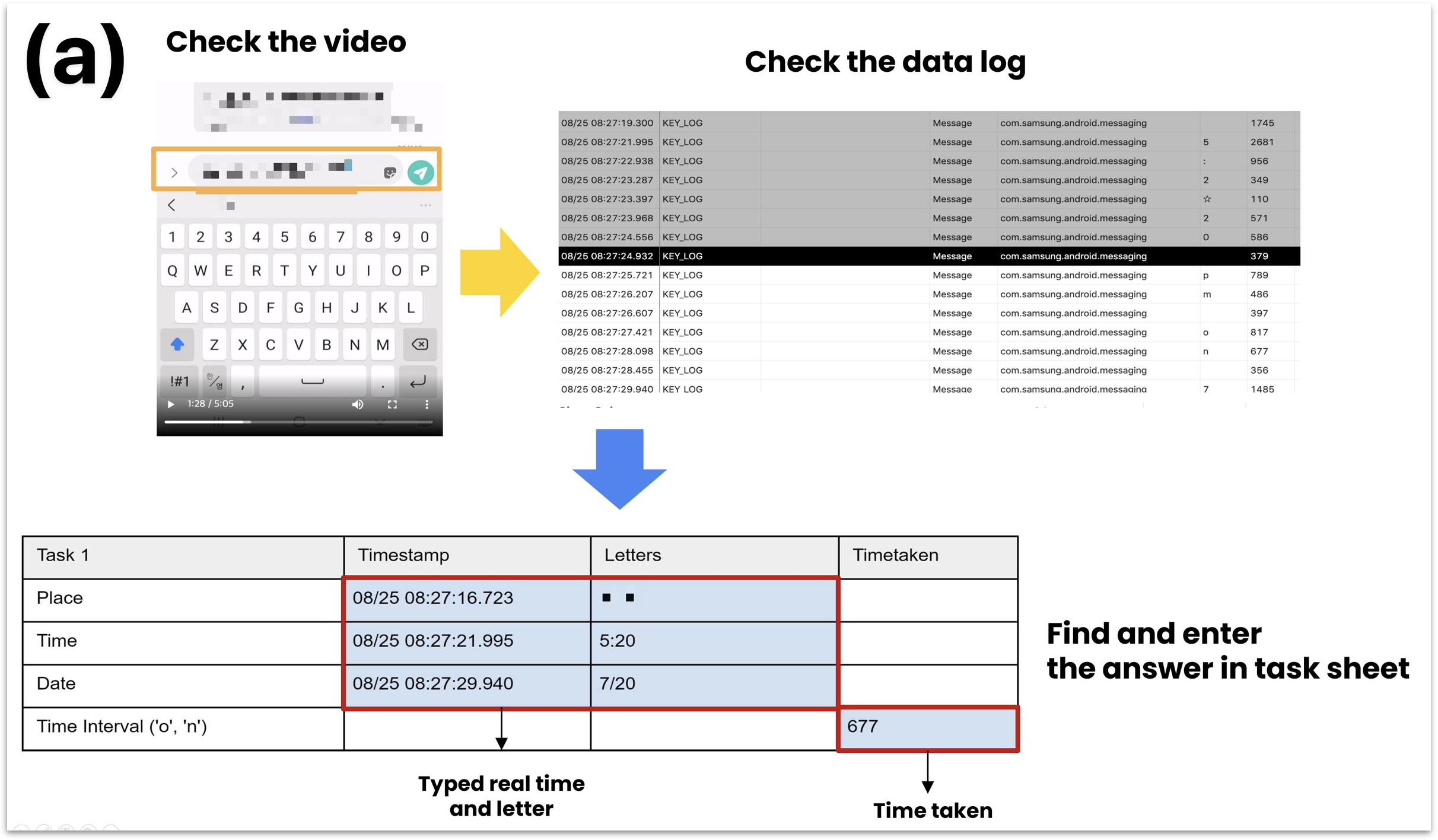}}
\subfigure{\includegraphics[width=0.48\textwidth]{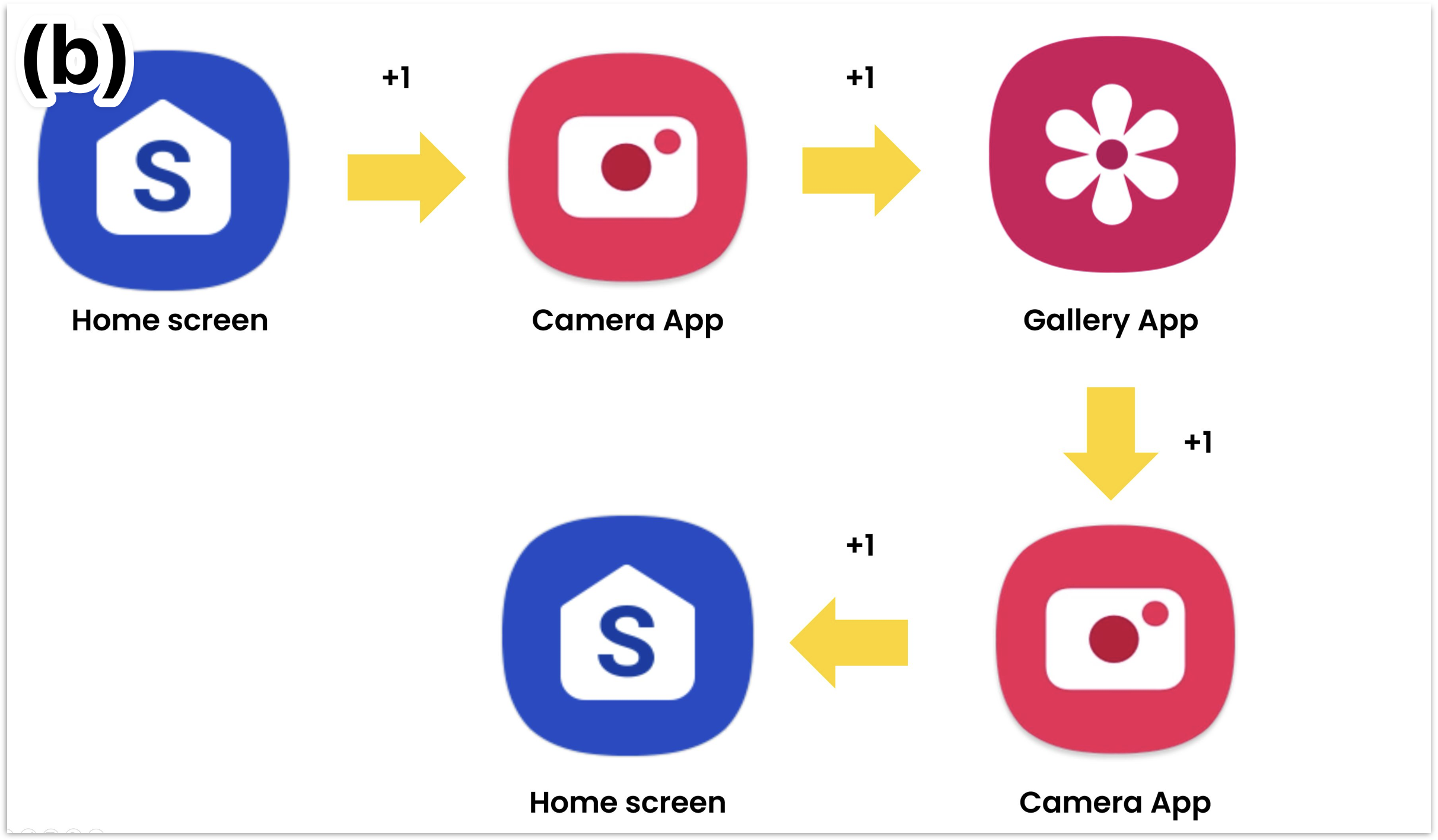}}
\subfigure{\includegraphics[width=0.48\textwidth]{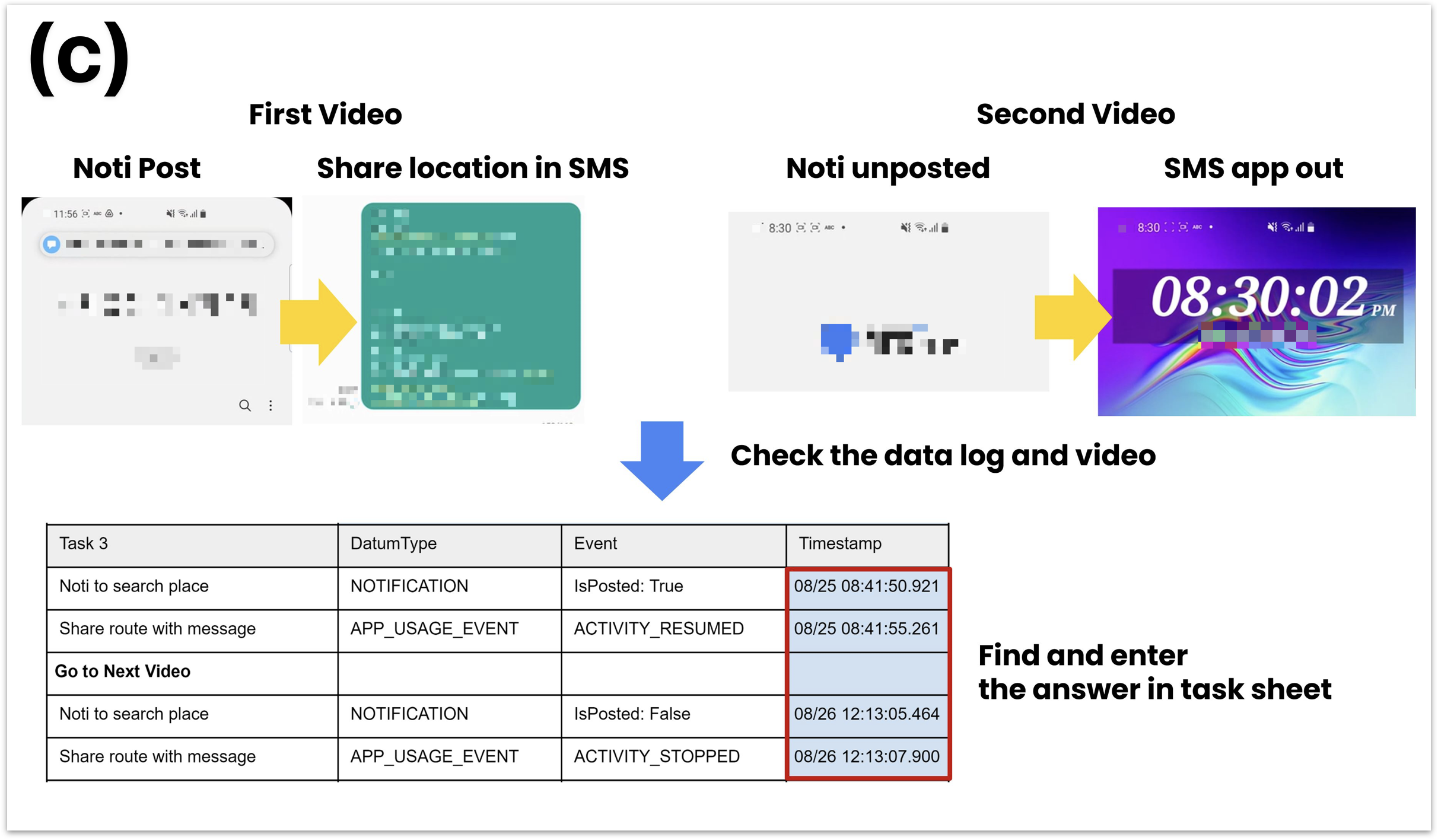}}
\subfigure{\includegraphics[width=0.48\textwidth]{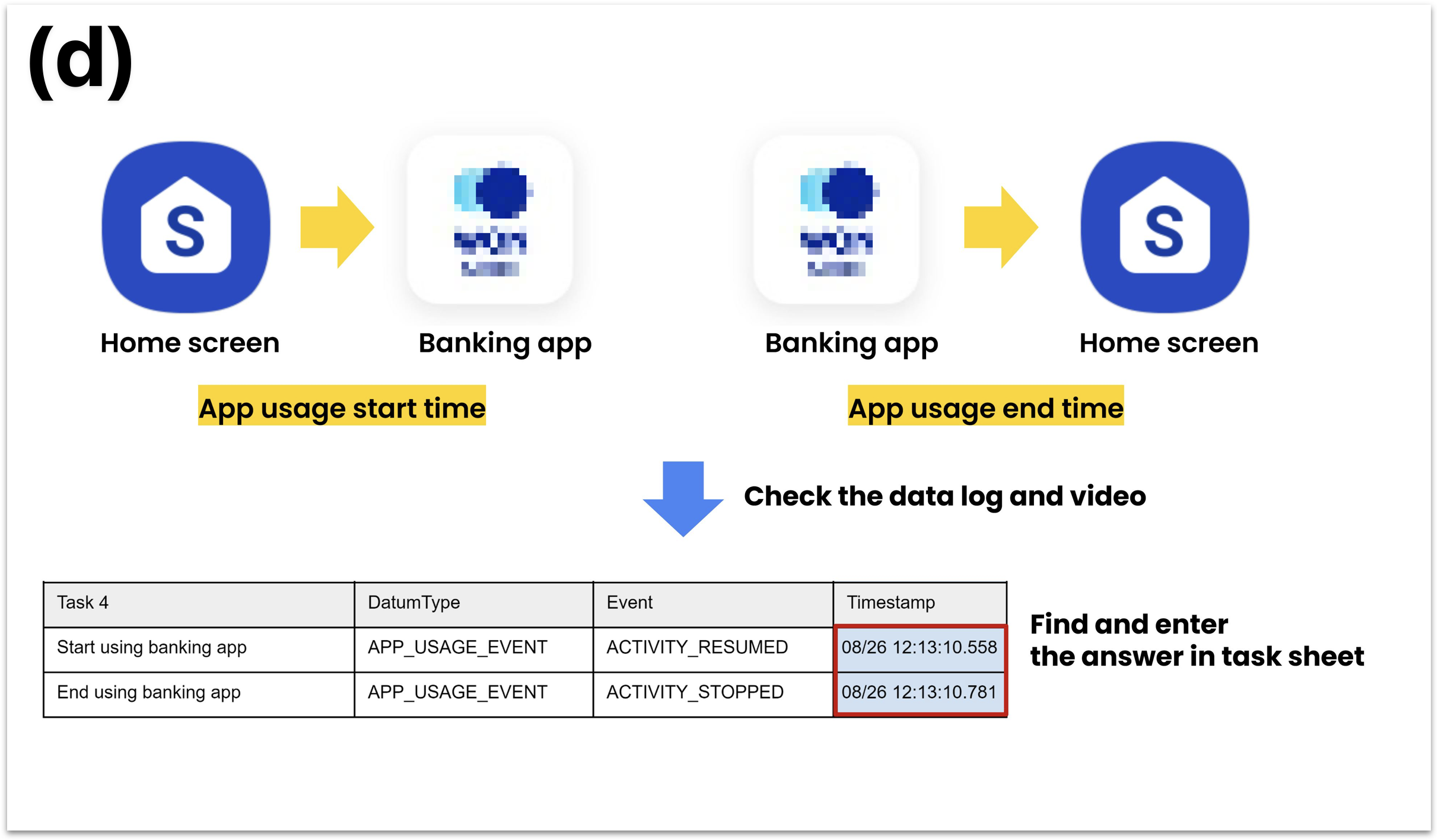}}
\caption{Task Description (a) App Keyboard typing analysis
 (b) The number of App transition analysis (c) App usage start/end time analysis in different time (d) App usage start/end time analysis in continuous time}
\label{tasks}
\end{figure}

Then, we explained the LV-Linker system and the tasks the participants should perform. Four tasks that participants should perform to analyze app usage data. Tasks are selected based on the app usage data analyzed in previous studies \cite{singh2017usage, shin2012understanding, qin2018deciphering, pielot2014situ}. In the first task, they should find typing information entered by text in the SMS app and find a time interval between two entered specific letters. In the second task, they should find the number of app transitions between Camera \& Gallery app. In the third task, there are two videos recorded at different times and log data which includes logs in two videos. In the first video, they find the time when the notification was posted and when the message was shared on Google Map. In the second video, they find the time when the notification was unposted and when the SMS app ended. In the last task, they find when an app started/resumed and ended/paused  in the banking app. The detailed description of each task is shown in Figure \ref{tasks}.

In the main experiment, participants were divided into two groups (group A and B) according to the order of use of the system (linked, unlinked) and performed the above four tasks. Group A used the unlinked system after using the linked system, and group B are in reverse order. The total experiment was 90 minutes long. In the first session, which is a practice session, participants learned how to use the system and basic knowledge of app usage data. We asked for the same four tasks for each session so that participants work on two iterations (practice, main). For quantitative evaluation, we measured the task completion time and the accuracy of task performance. Task completion time was measured every time participants finished each task, and when they finished one session, researchers recorded their time.

Last, the participants were asked to perform a post-survey. The post-survey asked questions about the helpfulness of the linked system for four tasks as 5 Likert scales from three perspectives: 1) correctness, 2) quickness, and 3) data understanding. We measured cognitive load using NASA Task Load Index (NASA-TLX) when analyzing app usage data with linked and unlinked systems, respectively: Mental/Physical/Temporal Demand, Effort, Performance, Frustration. Last, we asked for user feedback about the system.

\subsection{Result and Findings}
To determine statistical analysis methods, we examined the normality of Kolmogorov-Smirnov and Shapiro-Walk for quantitative and qualitative evaluation results. To analyze the result between the within-subject and the between-subject that is repeated differences for the results in the non-parametric statistics, the two-way mixed ANOVA repeated messages (RM) and turkey post hoc are used. In quantitative evaluation results, the task completion time of an experienced and a non-experienced person was not properly collected, so these two participants' data were excluded.

\begin{table}[]
\centering
\caption{Two-way mixed ANOVA RM results on task completion time for within linked/unlinked system and between expert/novice group or A/B group}
\label{anova}
\resizebox{\textwidth}{!}{%
\begin{tabular}{@{}llrrllrrl@{}}
\toprule
\textbf{Task Time} &
  \textbf{Source} &
  \multicolumn{1}{c}{\textbf{d.f}} &
  \multicolumn{1}{c}{\textbf{F value}} &
  \multicolumn{1}{c}{\textit{\textbf{P}}} &
  \textbf{Source} &
  \multicolumn{1}{c}{\textbf{d.f}} &
  \multicolumn{1}{c}{\textbf{F value}} &
  \multicolumn{1}{c}{\textit{\textbf{P}}} \\ \midrule\midrule
 &
L/U &
3 &
5.156 &
 0.01103* &
L/UL &
3 &
5.628 &
0.00789** \\
 &
E/N &
1 &
9.948 &
0.00614** &
  A/B &
  1 &
  2.612 &
  0.12558 \\
 &
  L/U * E/N &
  3 &
  0.467 &
  0.709 &
A/B * L/UL &
3 &
3.746 &
0.03266* \\
\multirow{-4}{*}{Overall Task} &
  error &
  16 &
  \multicolumn{1}{l}{} &
   &
  error &
  16 &
  \multicolumn{1}{l}{} &
   \\ \midrule
 &
  L/U &
  3 &
  1.525 &
  0.246 &
  L/UL &
  3 &
  1.822 &
  0.152 \\
 &
  E/N &
  1 &
  2.028 &
  0.174 &
  A/B &
  1 &
  2.946 &
  0.105 \\
 &
  L/U * E/N &
  3 &
  0.331 &
  0.803 &
  A/B * L/UL &
  3 &
  3.746 &
  0.143 \\
\multirow{-4}{*}{Task 1} &
  error &
  16 &
  \multicolumn{1}{l}{} &
   &
  error &
  16 &
  \multicolumn{1}{l}{} &
   \\ \midrule
 &
L/U &
3 &
6.388 &
0.00473** &
L/UL &
3 &
7.488 &
0.00237** \\
 &
E/N &
1 &
4.712 &
0.04534* &
  A/B &
  1 &
  0.329 &
  0.57445 \\
 &
  L/U * E/N &
  3 &
  0.521 &
  0.67369 &
A/B * L/UL &
3 &
3.262 &
0.04903* \\
\multirow{-4}{*}{Task 2} &
  error &
  16 &
  \multicolumn{1}{l}{} &
   &
  error &
  16 &
  \multicolumn{1}{l}{} &
   \\ \midrule
 &
L/U &
3 &
4.409 &
0.0193* &
L/UL &
3 &
3.612 &
0.0348* \\
 &
E/N &
1 &
8.319 &
0.0108* &
  A/B &
  1 &
  1.003 &
  0.3314 \\
 &
  L/U * E/N &
  3 &
  0.276 &
  0.8419 &
  A/B * L/UL &
  3 &
  0.343 &
  0.7944 \\
\multirow{-4}{*}{Task 3} &
  error &
  16 &
  \multicolumn{1}{l}{} &
   &
  error &
  16 &
  \multicolumn{1}{l}{} &
   \\ \midrule
 &
  L/U &
  3 &
  1.987 &
  0.157 &
  A/B Session &
  1 &
  1.009 &
  0.3301 \\
 &
  E/N &
  1 &
  1.043 &
  0.322 &
  L/UL &
  3 &
  2.711 &
  0.0797 \\
 &
  L/U * E/N &
  3 &
  0.259 &
  0.854 &
  A/B * L/UL &
  3 &
  2.432 &
  0.1028 \\
\multirow{-4}{*}{Task 4} &
  error &
  16 &
  \multicolumn{1}{l}{} &
   &
  error &
  16 &
  \multicolumn{1}{l}{} &
   \\ \bottomrule
\end{tabular}%
}
\end{table}
\subsubsection{Task Completion Time}
First, a two-way mixed ANOVA RM test was performed between expert/novice groups and on each of the five tasks (Task 1, 2, 3, 4, Overall task) completion times for the within four-session (practice and main session with linked/unlinked system). As shown in Table \ref{anova}, we found that there was a significant difference within four-session and between expert/novice groups in the total task, tasks 2, and 3 (p<0.05). However, we found that there were no significant interaction effects of the two factors.

Second, a two way mixed ANOVA RM test was performed for five task completion times between group A/B and within the four-session, respectively. As a result, we found that there was no significant difference between groups A/B in all five task completion times. As before, we analyzed that there was a significant difference between within-subjects in the total task, task 2, and 3 (\textit{p}<0.05) in Table \ref{anova}. However, we found that there was a significant interaction effect of the two factors which are the group A/B and session in the total task and task 2 (\textit{p}<0.05) in Table \ref{anova}.

In group A, the main session of task completion time was much shorter compared to the practice session. Accordingly, the linked system can help users to understand the log data in hard levels, so the learning effect was high when the users used the linked system first. The difference in task completion time within linked/unlinked systems and between expert/novice groups was significant in Task 2, 3, not Task 1, 4. Because Task 1, 4 is easier (see more video than log) than Task 2, 3. Therefore, the more difficult the task is, the more effective the linked system is than the unlinked system.

\subsection{Helpfulness \& Task Workload}
\begin{figure}
  \centering
  \includegraphics[width=\textwidth]{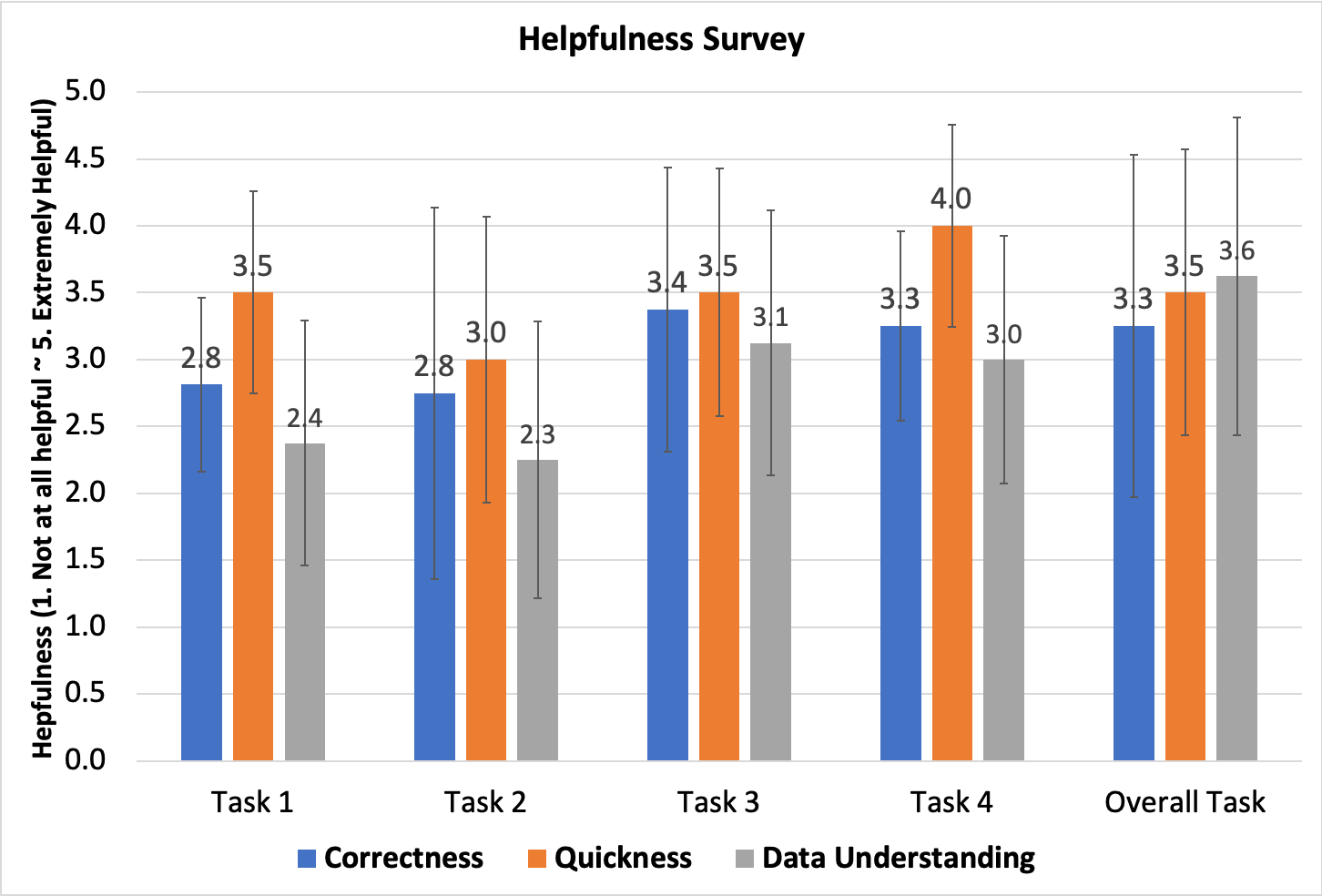}
  \caption{Helpfulness of Linked System for each Task Result,}
  \label{helpfulness}
\end{figure}

In the helpfulness survey, 75\% of respondents answered that the linked system was more helpful than the average in the overall task in terms of data understanding, correctness, and quickness compared to the unlinked system in Figure \ref{helpfulness}. In each data understanding, correctness, and quickness, they answered that the linked system was more helpful than the unlinked system in Tasks 3 and 4 in Figure \ref{helpfulness}. As a result of two-way within ANOVA (five tasks: Task 1, 2, 3, 4, Overall task;  survey items: data understanding, correctness, quickness), a significant difference only occurred for five task factors (\textit{p}<0.05). There was no interaction effect between the two factors. Furthermore, two-way mixed ANOVA analysis between the expert/novice group and within the five tasks, there was no significant difference, and no interaction effect as well. 

\begin{figure}
  \centering
  \includegraphics[width=\textwidth]{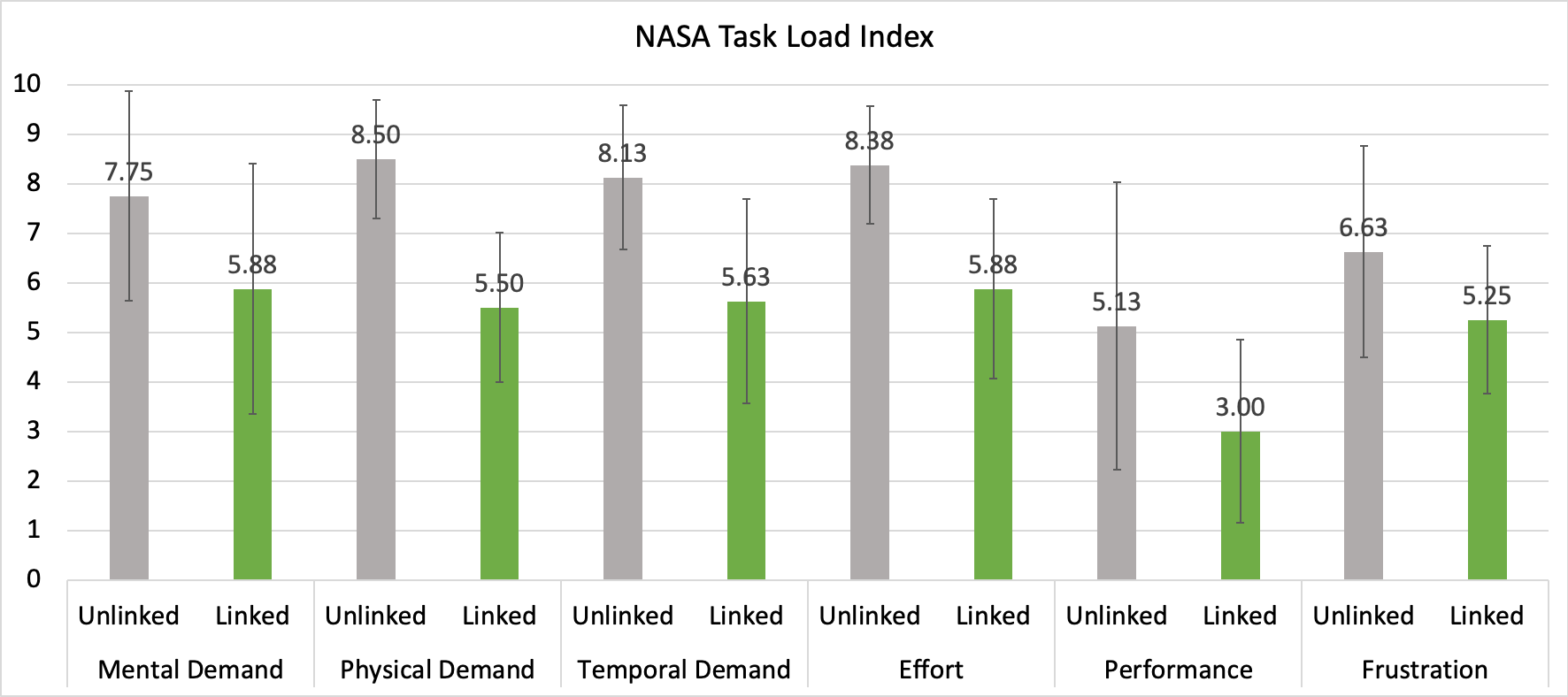}
  \caption{NASA Task Load Index Result}
  \label{nasa}
\end{figure}

In the NASA TLX survey, the average load reduction of the linked system compared to the unlinked system of NASA TLX among all participants in the experiment was performance (42\%), physical demand (35\%), temporal demand (31\%), effort (0.30\%), mental demand (24\%), and frustration (20\%) showed a reduction of more than 30\% in the four categories as shown in Figure \ref{nasa}. Users responded that the linked system reduced the user's load by 30\% or more compared to the unlinked system in the performance, physical/temporal demand, and effort items. Moreover, we found that there was a significant difference in all items except for the frustration item between linked/unlinked systems in NASA TLX (\textit{p}<0.05). We found that there was a significant difference between expert/novice groups in mental demand items by two-way mixed ANOVA RM for each survey item between expert/novice groups and within a linked/unlinked system in NASA TLX (p<0.05), but no interaction effect between the two factors. We found that the linked system had a lower load compared to the unlinked system in both expert and novice groups. In helpfulness, the linked system is to reduce the load of users in quickness, data understanding, and correctness in performing app usage analysis tasks.

\subsection{Subjective User Experience Reports}
We received short answers on the advantages \& disadvantages of linking log \& video and system improvements from all participants. Participants responded that linking was very helpful for quickness and data understanding. Non-experienced participants said that it was good to intuitively know which data is related to which activity. Experienced participants said that the location where the log starts can be immediately found through the video, so the action of scrolling to the corresponding part is omitted, which is likely to increase speed. However, some participants expressed concern about the system. They responded that if there is abnormal data, it is not sure which data (log or video) to believe. It can be more confusing than when using an unlinked system. In addition, they answered that it was difficult to accurately pause the specific point in the video because the event passed in an instant. As a system improvement, they responded that it would be good to add a function to automatically pause the video when certain log data is selected during video playback, and a function to label log event information on the video progress bar.

\section{CONCLUSION AND FUTURE WORK}
This study verified the effectiveness of app usage analysis through quantitative and qualitative evaluation results through LV-Linker, a system that proposed the linking effect of app usage log and video. The proposed LV-Linker system is expected to help analyze the app usage behavior pattern in various mobile and HCI studies using app usage logs and video data. In future work, our proposed system can link log data on the web or other smart devices (e.g., smartwatch) and video (display information) to analyze user interaction data.

\bibliographystyle{unsrt}
\bibliography{bibliography.bib}

\end{document}